\documentclass[]{vgtc}                          

\ifpdf
  \pdfoutput=1\relax                   
  \usepackage{graphicx}                
  \DeclareGraphicsExtensions{.pdf,.png,.jpg,.jpeg} 
\else
  \usepackage{graphicx}                
  \DeclareGraphicsExtensions{.eps}     
\fi%

\graphicspath{{figures/}{pictures/}{images/}{./}} 

\usepackage{subcaption}                 
\usepackage{microtype}                 
\PassOptionsToPackage{warn}{textcomp}  
\usepackage{textcomp}                  
\usepackage{mathptmx}                  
\usepackage{times}                     
\usepackage{cite}                      
\usepackage{tabu}                      
\usepackage{booktabs}                  
\newcommand{\ra}[1]{\renewcommand{\arraystretch}{#1}}
\usepackage[table,xcdraw]{xcolor}
\usepackage{hyperref}

\onlineid{0}

\vgtccategory{Research}

\usepackage{amsmath,amssymb,amsfonts}

\usepackage{algorithm}
\usepackage[noend]{algpseudocode}

\usepackage{xspace}
\usepackage{tikz}
\usepackage{pgfplots}
\pgfplotsset{compat=1.15}
\usepackage{pgfplotstable}

\usepackage[capitalize]{cleveref}

\renewcommand{\paragraph}[1]    {\medbreak\noindent {\bf #1}}       

\newcommand{\RR}{\mathbb{R}}

\newcommand{\graph}{\mathcal{G}}
\newcommand{\tree}{\mathcal{T}}

\newcommand{\vtkm}{\textrm{VTK-m}\xspace}
\newcommand{\nyx}{\textit{NYX}\xspace}
\newcommand{\isopres}{\textit{IP}\xspace}
\newcommand{\magrec}{\textit{MAG}\xspace}
\newcommand{\richt}{\textit{RM}\xspace}
\newcommand{\cham}{\textit{CHAM}\xspace}
\newcommand{\truss}{\textit{TRUSS}\xspace}
\newcommand{\woodbranch}{\textit{WOOD}\xspace}
\newcommand{\paw}{\textit{PAW}\xspace}
\newcommand{\jicf}{\textit{JICF-Q}\xspace}

\pgfplotsset{tmtomp/.style=
    {mark=*,
     mark options={fill=green},
     color=green,
    }}

\pgfplotsset{tmtompip/.style=
    {mark=*,
     mark options={fill=red},
     color=red,
    }}

\pgfplotsset{tmtompmr/.style=
    {mark=diamond*,
     mark options={fill=red},
     color=red,
    }}

\pgfplotsset{tmtomprm/.style=
    {mark=triangle*,
     mark options={fill=red},
     color=red,
    }}

\pgfplotsset{tmtompcosmo/.style=
    {mark=pentagon*,
     mark options={fill=red},
     color=red,
    }}

\pgfplotsset{tmtgpu/.style=
    {mark=diamond*,
     mark options={fill=blue},
     color=blue,
    }}

\pgfplotsset{vtkomp/.style=
    {mark=triangle*,
     mark options={fill=brown},
     color=brown,
    }}

\pgfplotsset{vtkgpu/.style=
    {mark=pentagon*,
     mark options={fill=red},
     color=red,
    }}

\pgfplotsset{vtkompip/.style=
    {mark=*,
     mark options={fill=blue},
     color=blue,
    }}

\pgfplotsset{vtkompmr/.style=
    {mark=diamond*,
     mark options={fill=blue},
     color=blue,
    }}

\pgfplotsset{vtkomprm/.style=
    {mark=triangle*,
     mark options={fill=blue},
     color=blue,
    }}

\pgfplotsset{vtkompcosmo/.style=
    {mark=pentagon*,
     mark options={fill=blue},
     color=blue,
    }}

\pgfplotsset{nesoi/.style=
    {mark=square*,
     mark options={fill=cyan},
     color=cyan,
    }}

\title{Fast Merge Tree Computation via SYCL}
\author{
      Arnur Nigmetov\thanks{e-mail:anigmetov@lbl.gov} \and
      Dmitriy Morozov\thanks{e-mail:dmorozov@lbl.gov}}
\affiliation{\scriptsize Lawrence Berkeley National Laboratory}

\CCScatlist{
    \CCScatTwelve{Computing methodologies}{Massively parallel algorithms}{}{};
    \CCScatTwelve{Theory of computation}{Computational geometry}{}{};
}

\abstract{
A merge tree is a topological descriptor of a real-valued function.
Merge trees are used in visualization and topological data analysis,
either directly or as a means to another end: computing a 0-dimensional persistence diagram,
identifying connected components, performing topological simplification, etc.
 
Scientific computing relies more and more on GPUs to achieve fast, scalable
computation. For efficiency, data analysis should take place at the same
location as the main computation, which motivates interest in parallel
algorithms and portable software for merge trees that can run not only on a CPU, but also
on a GPU, or other types of accelerators.
The SYCL standard defines a programming model that allows the same code, written
in standard C++, to compile targets for multiple parallel backends (CPUs via
OpenMP or TBB, NVIDIA GPUs via CUDA, AMD GPUs via ROCm, Intel GPUs via Level Zero, FPGAs).
In this paper, we adapt the triplet merge tree algorithm to SYCL
and compare our implementation with the \vtkm implementation, which is the only
other implementation of merge trees for GPUs that we know of.
}


\date{\today}

\begin{document}

\maketitle

\section{Introduction}

Let $f$ be a continuous scalar function defined on some domain $D$ in $\RR^n$. Let us consider connected components of its sublevel sets
$f^{-1}(-\infty, t]$, as $t$ changes from $-\infty$ to $\infty$. 
Each time we pass a local minimum, a new connected component appears; existing connected components 
merge together as we pass a saddle. We say that two points $x, y \in D$ are equivalent, if $f(x) = f(y)$, and both $x$ and $y$
belong to the same connected component of $f^{-1}(-\infty, f(x)]$. Taking a
quotient of the domain by this equivalence relation, we obtain a tree, 
called a \emph{merge tree}. Each branch of the tree corresponds to a pair of critical points $(a, b)$, where $a$ is a local minimum and $b$ is a saddle.
Short branches, i.e., branches such that $\lvert f(a) -f(b) \rvert$  is small, can be intuitively interpreted as topological noise, 
small `wrinkles' on the graph of $f$. Longer branches encode more significant features of the function. Merge tree is an example of a topological descriptor.

Another topological descriptor is a \emph{persistence diagram} --- in our case
specifically, a zero-dimensional persistence diagram.
It is a multi-set of points $(f(a), f(b)) \in \RR^2$, one point per branch $(a, b)$ of the merge tree.
It can be viewed as `bag of features': compared to the merge tree, we lose the
information about how the branches merge together and keep only their endpoints.
Persistence diagrams are one of the main tools of topological data analysis.
Since they are not the main topic of this article, we refer an interested reader to
the textbook~\cite{edelsbrunner2022computational} for a proper definition of persistence diagrams in all
dimensions.

The last topological descriptor that we want to mention is a \emph{contour
tree}, which captures the structure of level sets on simply connected domains,
i.e., domains where every loop can be contracted to a point.
Formally, we declare two points $x, y \in D$ equivalent, if $f(x) = f(y)$, and $x$ and $y$ are in the same connected component of
$f^{-1}(f(x))$. We get the contour tree by taking the quotient of $D$ by this equivalence relation.
Contour tree is a well-known tool in visualization community, since
it allows to graphically present the structure of high-dimensional scalar fields.

In practice, we do not work with continuous functions on continuous domains. Typically, a scalar function
is given only on the points of a grid. We can connect neighboring grid points, obtaining
a graph with the function defined on its vertices. This is a standard setting, 
in which all the aforementioned topological descriptors are computed in practice .

We are interested in merge trees for the following reasons:
\begin{itemize}
    \item There are many applications of merge trees per se, including halo finding
          in cosmology~\cite{friesen2016situ}, finding atmospheric rivers in climatology~\cite{inatsu2017cyclone},
          tracking features in combustion simulations~\cite{bremer2010interactive,mascarenhas2011topological,bennett2011feature},
          among many others.
    \item One of the standard algorithms for computing contour trees,
          proposed by Carr et al.~\cite{carr2003computing}, computes the contour tree
          of $f$ from the merge trees of the functions $f$ and $-f$ (the authors
          call them the \emph{join tree} and the \emph{split tree}).
    \item The connection between merge trees and persistence diagrams is not just a shortcut
        for defining the diagram in dimension $0$, but it is a practically efficient
        way to compute it. Persistence diagrams are used,
        e.g., in neuroscience~\cite{petri2014homological}, material science~\cite{buchet2018persistent}, 
        chemistry~\cite{krishnapriyan2020topological}, etc.
        We can also refer the reader to the database of articles
        with real-world application of TDA~\cite{giuntidb} for more applications.
        Note that for large inputs the zero-dimensional persistence diagram is
        the only computable option.
\end{itemize}



Often in these applications the input function is a result
of a numerical simulation on a supercomputer.
As the supercomputer architectures evolve and most of the computation is
offloaded to fast accelerators, most commonly GPUs, data analysis needs to keep
up with these changes and adapt the algorithms to run on the same devices that
contain the data. There are many reasons for this: computational efficiency,
minimizing data movement (both to reduce runtime and for energy efficiency), the
need for in situ analysis to provide rapid feedback to the simulation, among
many others.
All these factors motivate developing efficient, \emph{portable} algorithms that
can run on as wide of a range of the devices as the simulations themselves.
In addition, the emerging interest in using topological information in machine
learning, especially, using persistence diagrams to guide the learning
process~\cite{topological-regularizer,differential-calculus-barcodes,pso},
motivates developing efficient implementations of topological algorithms that
can execute on the same devices as the machine learning pipelines, i.e.,
predominantly on GPUs.


\paragraph{Related work.}
In the serial setting, merge trees can be computed using a variation
of the classical Kruskal's algorithm. There are several papers proposing different
shared-memory parallel algorithms
\cite{pascucci2004parallel, bremer2010interactive, gueunet2017task}.
Carr et al.~\cite{parpeak} proposed a parallel algorithm for computing merge
trees, implemented in \vtkm
with OpenMP and Thrust (CUDA). We recall more details about this algorithm
in the next section, because we use that implementation as a baseline for comparison.
Smirnov and  Morozov~\cite{tmt} developed the concept of
a triplet merge tree, with a simple path-merging algorithm to compute it.
The main advantage of this algorithm is that it is lock-free, using the
compare-and-swap idiom.
Since this is the algorithm that we implement using SYCL
in this paper, we also review it in the next section.


\paragraph{}
Our contributions are as follows:
\begin{itemize}
    \item We adapt the triplet merge algorithm~\cite{tmt} to SYCL, a programming model
        targeting multiple processing units (including CPUs, GPUs, and other
        accelerators).
    \item We experimentally evaluate our implementation, comparing it to the
        \vtkm implementation of merge trees, which is the only implementation
        that we know of that runs on GPUs (and CPUs).
\end{itemize}

\section{Background}
\subsection{Merge Trees}
Let $f \colon X \to \RR$ be a function on a space $X$. The merge tree of $f$
is the quotient space $\tree_f = X/~$, where we identify two points
$x_1$ and $x_2$ if and only if $f(x_1) = f(x_2) =: y$, and $x_1$
and $x_2$ belong to the same connected component of the sublevel set $f^{-1}(-\infty, y]$.
In our setting, the space is a graph $\graph = (V, E)$, and the function
is defined only on the vertices, $f \colon V \to \RR$.
We extend $f$ to edges by linear interpolation and write $f \colon \graph \to \RR$.
For $c \in \RR$, we use $G_c$ to denote the sublevel set:
$G_c := \{ v \in V \mid f(v) \leq c \}$.
The \textit{merge tree} of $f$ is a graph $\tree_f$ with the same vertex set $V$.
Assuming that $f(v_1) \leq f(v_2)$,
there is an edge $v_1 v_2$ in $\tree_f$ if and only if
there does not exist a vertex $u$ such that $f(v_1) \leq f(u) \leq f(v_2)$
and $v_1$ and $u$ are in the same component of $G_{f(u)}$.

Merge trees can be computed in $O(m \log n)$ time, on a graph with $n$ vertices
and $m$ edges, using the union--find (disjoint sets) data structure.
First, one sorts the vertices by values of the function.
Then the algorithm goes over the sorted vertices
and uses the union--find data structure to determine
which of the three alternatives occurs:
(1) the current vertex is a local minimum, so it creates a new connected
component in the sublevel set; (2) the current vertex belongs to exactly
one of the components that were present at the previous vertex; (3)
multiple connected components are merged together at the current vertex.
The requirement to process the vertices in sorted order makes it difficult to
parallelize this algorithm.

\paragraph{Peak pruning.}
Carr et al. \cite{parpeak} compute the merge tree
in parallel avoiding global sorting.
For each vertex $u$ that is not a local minimum,
they pick an arbitrary edge $uv$
such that $f(v) < f(u)$, a completely independent and thus easily parallelizable
operation.
They follow the selected edges $f(u) > f(v) > f(w) > \cdots > f(z)$
to build a path from every vertex $u$ to a local minimum $z$;
they say that $u$ is \textit{assigned} to $z$.
Among all vertices assigned to the same local minimum $z$ the authors identify
those that can be saddles; such a saddle candidate must have two neighbors
below it that are assigned to different local minima. They prove that the lowest saddle candidate
assigned to $z$ is exactly the saddle paired with $z$. That is, the lowest
saddle candidate $s$ is the saddle at which the sublevel set component
born at $z$ merges into another one. This immediately gives
the pairing of the critical points and the branch structure of the merge tree.
The remaining part of their algorithm takes care of regular vertices
(those vertices that have degree two in the merge tree).
Some of those are readily available: all vertices that are assigned to $z$ and
are below $s$ must be on that branch.
The authors remove all these vertices and repeat the same procedure
(note that some of the saddles became minima after removal).
This algorithm is implemented in VTK-m~\cite{vtkm}, and we use it as a baseline for
comparison.

\paragraph{Triplet merge trees.}
A different parallel algorithm is described in \cite{tmt}.
Its main idea is to replace the standard merge tree, described
above, by its dual tree of branches. Each branch is a path that tracks when a component is
created and when it is merged with an older component.
It is represented as a \emph{triplet} of vertices $u$, $s$, $v$, such that $f(v)
< f(u) \leq f(s)$ and $u$ and $v$ are in the same connected component of the
sublevel set $\graph_{f(s)}$.
The triplet means that a branch created by vertex $u$ merges with a branch
created by vertex $v$ at vertex $s$.
Additionally, if $u$ is a global minimum of $f$ in its connected component of
the graph, then, by convention, $(u, u, u)$  is a triplet.
It is convenient to interpret each triplet as a directed edge $(u,v)$ with a label $s$.
Vertices of degree two in a standard merge tree (i.e., the inner vertices
of a branch) become leaves in the triplet merge tree: they are represented by
triplets $(u, u, v)$.
To get a unique representation of a merge tree as a collection of triplets,
we need two conditions: (1) each vertex $u$ appears exactly once
as a first vertex of some triplet; (2) for each $(u, s, v)$, vertex $v$ is the deepest
vertex in $f^{-1}(-\infty, f(s)]$. The first condition is called \textit{normalization},
the second one \textit{minimality}.

In \cite{tmt} the authors explain the details of a lock-free algorithm
to compute the triplet representation of $\tree$.
We reproduce the algorithm  here for convenience, see \cref{alg:comp_mt}.
Since we want to maintain the normalization condition from the start,
we represent the tree as a map $T$, where the first vertex $u$ of a triplet is the key;
the entries are pairs $T[u] = (s, v)$ --- the second and the third
vertices of the triplet. In the original CPU implementation, $T[u]$ is
a pointer to a pair, not a pair itself. We ignore that in the pseudocode, but
return to this issue in \cref{subsec:gpu_changes}, where we explain the
changes needed for the GPU.

In the first loop, we initialize the tree with triplets $(u, u, u)$ (which corresponds to the case of no edges, $E = \emptyset$).
Then the algorithm works in two phases: merge and repair.

The merge phase (lines 4--8) processes the edges $E$ in parallel: for each edge $(u, v)$ it starts
with a normalized merge tree on a graph missing this edge and changes it to
incorporate $(u, v)$.  The proof of correctness can be found in \cite{tmt}.
The updates are synchronized using compare-and-swap operations (\cref{alg:merge}, line 14),
which guarantees correctness. Logically, compare-and-swap (CAS, \cref{alg:cas})
checks if the variable $v$ that we want to modify has the value that we expect (usually
this means that another thread has not modified it since we read its value),
and, only in that case, changes it to the desired value. It returns the result
of the comparison. All these operations are performed atomically,
as a single transaction.
If the  comparison failed, we simply start from scratch (line 15).

The repair phase fixes the minimality condition that may be violated in the
merge phase.  Let $u$ be a vertex, $a$ a real number. We call vertex $v$
with the smallest function value in the connected component of $u$
in $f^{-1}(-\infty, a]$ the \emph{representative}
of $u$ at level $a$. If we view triplets $(u, s, v)$ as directed edges
from $u$ to $v$, then, to find a representative of $u$, we just need to follow these edges until we reach
the deepest vertex. This is exactly the role of \cref{alg:repr}.
To ensure minimality, in \cref{alg:repair} we simply replace triplet $(u, s, v)$ with the
triplet $(u, s, v')$, where $v'$ is the representative of $u$.

\begin{algorithm}
\begin{algorithmic}[1]
\Function{ComputeMergeTree}{$G$}
    \ForAll {vertex $u \in G$}\textbf{ in parallel}
        \State {$T[u] \gets (u, u)$}
   \EndFor
    \ForAll {edge $(u, v) \in G$}\textbf{ in parallel}
        \If {$f(v) < f(u)$}
            \State \Call {Merge}{$T$, $u$, $u$, $v$}
        \Else
            \State \Call {Merge}{$T$, $v$, $v$, $u$}
        \EndIf
   \EndFor
    \ForAll {vertex $u \in G$}\textbf{ in parallel}
        \State \Call {Repair}{$u$}
   \EndFor
   \State {\Return $T$}
\EndFunction
\end{algorithmic}
\caption{Triplet Merge Tree Computation.}
\label{alg:comp_mt}
\end{algorithm}

\begin{algorithm}
\begin{algorithmic}[1]
\Function{CAS}{$v$, expected, desired}
    \If {$v = \mbox{expected}$}
        \State $v \gets \mbox{desired}$
        \State \Return \textbf{True}
    \Else
        \State \Return \textbf{False}
    \EndIf
\EndFunction
\end{algorithmic}
\caption{Compare-and-Swap.}
\label{alg:cas}
\end{algorithm}

\begin{algorithm}
\begin{algorithmic}[1]
\Function{Merge}{$T$, $u$, $s$, $v$}
    \State {$(s_u, u') \gets T[u]$}
    \If {$f(s_u) < f(s)$}
        \State \Return \Call {Merge}{$T$, $u'$, $s$, $v$}
    \EndIf
    \State {$(s_v, v') \gets T[v]$}
    \If {$f(s_v) < f(s)$}
        \State \Return \Call {Merge}{$T$, $u$, $s$, $v'$}
    \EndIf
    \If {$u = v$}
        \State \Return
    \EndIf
    \If {$f(v) < f(u)$}
        \State \Call {swap}{$(u, s_u, u')$, $(v, s_v, v')$}
    \EndIf
    \If {\Call{CAS}{$T[v]$, $(s_v, v')$, $(s, u)$}}
        \State \Call {Merge}{$T$, $u$, $s_v$, $v'$}
    \Else
        \State \Call {Merge}{$T$, $u$, $s$, $v$}
    \EndIf
   \State {\Return $T$}
\EndFunction
\end{algorithmic}
\caption{Parallel Merge.}
\label{alg:merge}
\end{algorithm}

\begin{algorithm}
\begin{algorithmic}[1]
\Function{Representative}{$T$, $u$, $a$}
    \State {$(s, v) \gets T[u]$}
    \While { $f(s) \leq a$ and $s \neq v$}
        \State $u \gets v$
        \State $(s, v) \gets T[u]$
    \EndWhile
    \State \Return $v$
\EndFunction
\end{algorithmic}
\caption{Representative in Triplet Merge Tree.}
\label{alg:repr}
\end{algorithm}

\begin{algorithm}
\begin{algorithmic}[1]
\Function{Repair}{$T$, $u$}
    \State {$(s, v) \gets T[u]$}
    \State {$v' \gets $ \Call {Representative}{$T$, $u$, $f(s)$}}
    \If {$u \neq v' $}
        \State $T[u] \gets (s, v')$
    \EndIf
    \State \Return $T[u]$
\EndFunction
\end{algorithmic}
\caption{Repair Triplet Merge Tree.}
\label{alg:repair}
\end{algorithm}

\section{TMT-SYCL}

\subsection{SYCL}
SYCL is a programming model for writing heterogeneous
parallel programs. It was originally designed to provide
a layer of abstraction over OpenCL, but has since evolved into
an independent standard, with several independent implementations, not
bound to a particular parallel device.
The main feature of SYCL is that it uses a single-source code written in
standard C++.
The code for parallel
execution on device is written in the same file
with the host code, and takes advantage of the standard language constructs.
This single source is then processed in multiple passes of
a compiler (or different compilers for host
and device). Different implementations of SYCL target CPUs via OpenMP or TBB, NVIDIA
GPUs via CUDA, AMD GPUs via ROCm, Intel GPUs via Level Zero, Intel and Xilinx
FPGAs.  We use an implementation called hipSYCL~\cite{hipSYCL}.  It supports OpenMP, NVIDIA
CUDA and AMD (ROCm) backends.



\subsection{Changes needed for GPU}
\label{subsec:gpu_changes}
One limitation of SYCL is lack of support for dynamic memory allocation.
On-device buffers are declared, with specific size, before the execution of the
device code that uses them.
In the CPU implementation\footnote{Available publicly in Reeber,
\url{github.com/mrzv/reeber}} of \cite{tmt}, the tree is represented as a map
with vertices
as keys and pointers as values. The pointer corresponding
to vertex $u$ refers to an object that stores the two
vertices $s$ and $v$.
It also stores a vector of degree-2 vertices.
Note that in the CPU implementation vertices were represented
as double-word variables, which means that we cannot use compare-and-swap
to atomically update the pair of values $(s, v)$. Normal CAS
operates on single words, and double word CAS, available on \texttt{x86-64},
can update 2 contiguous words in memory atomically,
but we would need to update 4 words. This is the
reason for $T[u]$ being a pointer to a pair.

Although it is possible to implement a dynamic memory allocator on top of the
static buffers (and then implement a hash map on top), we choose a simpler
route.
The triplet $(u, s, v)$ is represented as an array of pairs: $T[u] = (s, v)$.
Another limitation of SYCL is that it does not provide a double-word
compare-and-swap operation, which the triplet merge algorithm requires to
atomically update the pair $(s,v)$. We choose a simple solution: we limit the
vertex identifier to a 32-bit integer. This way a pair can be packed into a
64-bit integer and a regular compare-and-swap operation, which is supported by
SYCL, suffices. This limitation is minor, given the current memory constraints
on GPUs: the data and the tree together for $2^{32}$ vertices, assuming 32-bit
floating point and 32-bit integers for the tree, would require 48 GiB on the device.





\section{Experiments}
\paragraph{Datasets.} We use the following datasets.
All of them, except \nyx, were downloaded from P.\ Klacansky's
database \cite{klacdb}.
We used averaging and interpolation to downsample the higher-resolution datasets.
\begin{itemize}
    \item Cosmology (\nyx) \cite{nyx}. This is a snapshot of the dark matter density
        from a Nyx simulation (for redshift $z=2$).
    \item Magnetic reconnection (\magrec) \cite{magrec}. This phenomenon is observed in plasma;
        the lines of magnetic field change their connectivity. The dataset is a single step
        of a simulation.
    \item Isotropic pressure (\isopres) \cite{ip}. The function is
        the pressure field of a direct numerical simulation of forced isotropic turbulence.
    \item Entropy field of Richtmyer--Meshkov instability (\richt) \cite{richtmeshk}. This is an
        event observed when two fluids of different density are intensively mixed with other
        by the impact of a shock wave.
    \item Three CT scans. \cham is a CT scan of a chameleon, scanned by DigiMorph.
          \woodbranch is a CT scan of a wood branch  by the Computer-Assisted Paleoanthropology group and 
        the Visualization and MultiMedia Lab at University of Zurich.
        \paw is a scan of a \textit{Pawpawsaurus Campbelli}.
        These datasets have rich topology, if one considers the evolution of superlevel sets, 
        because many connected components emerge at high density.
    \item \truss is a simulated CT scan of an $8\times8\times8$ octet truss \cite{synthetic_truss_with_five_defects}.
        This dataset has a specifically regular structure.
    \item \jicf is a $Q$ criterion of a jet in cross-flow \cite{jicf_q}. If $\vec{v}$ is a velocity vector field of fluid,
        then its gradient $\nabla v$ is a $3\times 3$ tensor. It can be decomposed into the symmetric part $S$ 
        and anti-symmetric part $\Omega$: $\nabla \vec{v} = S + \Omega$. $Q$ criterion is defined as $\frac{1}{2}(\| \Omega \|^2 - \| S \| ^2)$
        and is used to detect vortices.  Most of the topological evolution of this dataset is localized in a small part of the domain.
\end{itemize}

\paragraph{Setup.} Experiments were performed on a computer with 
Intel(R) Xeon(R) Gold 6230 CPU (20 physical cores),
NVIDIA GeForce RTX 2080 Ti GPU, running Arch Linux.
For comparison, we used \vtkm code for computing the contour tree.
Since \vtkm computes contour trees from two merge trees, 
following the algorithm of Carr et al.~\cite{carr2003computing},
we measured the running time
of both computations separately.
To this end, we modified the code of \vtkm,
timing both calls of the merge tree computation
routine separately, to make the comparison fair by excluding
the overhead to combine them into the contour tree.\footnote{It is difficult to measure the time it
takes to transfer data to GPU and back from it. However, the output of
\texttt{nvprof} shows that the transfer takes a relatively small time for both
\vtkm and tmt-sycl codes, about 10\% to 15\% of the total execution.}
However, in all these datasets the merge tree
of $-f$ carries more information
than the tree of $f$ (one can verify that by visually exploring
the function for different threshold values using a tool at \cite{klacdb}).
For example, for the cosmological dataset,
the branches of $\tree_{-f}$ are born at the local maxima of $f$,
which capture \textit{halos}, clusters of high density, while $\tree_f$ consists
of a single path.
Therefore, we report the results for $-f$ only.
Using the terminology of \cite{parpeak}, we are only interested in the \textit{split tree}.
Each experiment was run 5 times, and we report the average timing.
The running time is stable across the experiments,
showing only minor fluctuations of 5\% maximum.


%

\begin{figure}[t]
\centering
\pgfplotstableread[col sep = semicolon]{data/z2/z2_tmt_sycl_gpu.txt}\cosmotmtgpu
\pgfplotstableread[col sep = semicolon]{data/z2/z2_tmt_sycl_openmp.txt}\cosmotmtopenmp
\pgfplotstableread[col sep = semicolon]{data/z2/z2_vtk_gpu.txt}\cosmovtkgpu
\pgfplotstableread[col sep = semicolon]{data/z2/z2_vtk_openmp.txt}\cosmovtkopenmp
\begin{tikzpicture}
\begin{axis}[
        title={Cosmology, $z = 2$},
                ymode=log,
                xmode=log,
                height=2.5in,
                width=.45\textwidth,
                ymax=300,
                log ticks with fixed point,
                xtick={256,512,768,1024},
                xticklabels={$256^3$,$512^3$,$768^3$,$1024^3$},
                xlabel={Size},
                ylabel={Seconds},
                nodes near coords,      
                point meta=rawy,        
                legend pos=north west,
                legend style={font=\footnotesize},
                legend cell align={left},
                legend columns=1
            ]
    \addplot    +[vtkomp] table[x=lin_size, y=elapsed] \cosmovtkopenmp; \addlegendentry{VTK-m, OpenMP}
    \addplot    +[vtkgpu] table[x=lin_size, y=elapsed] \cosmovtkgpu; \addlegendentry{VTK-m, GPU}
    \addplot    +[tmtomp] table[x=lin_size, y=elapsed_total] \cosmotmtopenmp; \addlegendentry{tmt-sycl, OpenMP}
    \addplot    +[tmtgpu] table[x=lin_size, y=elapsed_total] \cosmotmtgpu; \addlegendentry{tmt-sycl, GPU}
\end{axis}
\end{tikzpicture}
    \caption{Running time comparison of tmt-sycl and \vtkm, on \nyx dataset,
             varying input size.}
\label{fig:cosmo_z2_comp}
\end{figure}
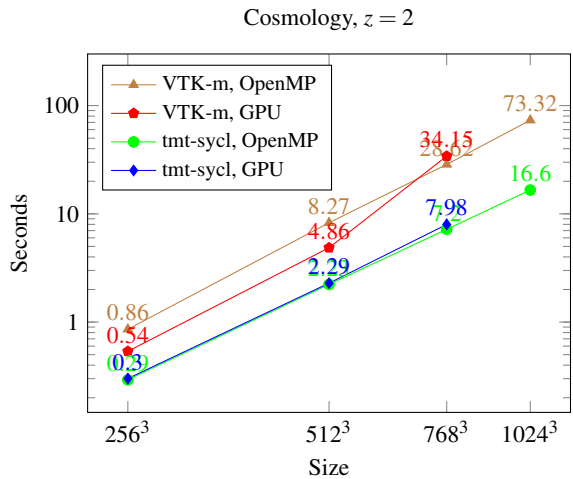

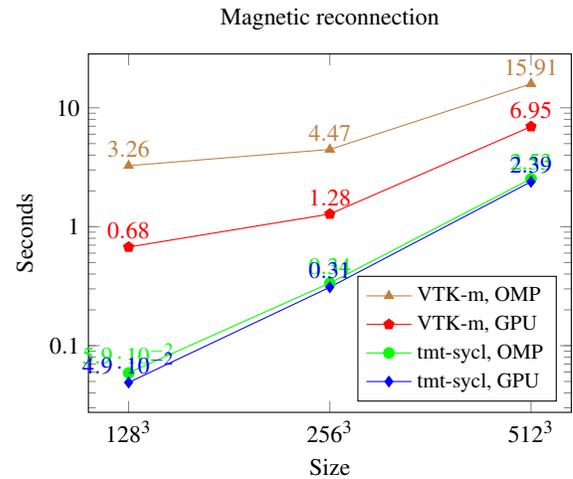
\begin{figure}[t]
\centering
\pgfplotstableread[col sep = semicolon]{data/magnetic_reconnection/mr_tmt_sycl_gpu.txt}\mrtmtgpu
\pgfplotstableread[col sep = semicolon]{data/magnetic_reconnection/mr_tmt_sycl_openmp.txt}\mrtmtopenmp
\pgfplotstableread[col sep = semicolon]{data/magnetic_reconnection/mr_vtk_gpu.txt}\mrvtkgpu
\pgfplotstableread[col sep = semicolon]{data/magnetic_reconnection/mr_vtk_openmp.txt}\mrvtkopenmp
\begin{tikzpicture}
\begin{axis}[
        title={Magnetic reconnection},
                ymode=log,
                xmode=log,
                height=2.5in,
                width=.45\textwidth,
                log ticks with fixed point,
                xtick={128,256,512},
                xticklabels={$128^3$, $256^3$,$512^3$},
                xlabel={Size},
                ylabel={Seconds},
                nodes near coords,      
                point meta=rawy,        
                legend pos=south east,
                legend style={font=\footnotesize},
                legend cell align={left},
                legend columns=1,
            ]
    \addplot    +[vtkomp] table[x=lin_size, y=elapsed] \mrvtkopenmp; \addlegendentry{VTK-m, OMP}
    \addplot    +[vtkgpu] table[x=lin_size, y=elapsed] \mrvtkgpu; \addlegendentry{VTK-m, GPU}
    \addplot    +[tmtomp] table[x=lin_size, y=elapsed_total] \mrtmtopenmp; \addlegendentry{tmt-sycl, OMP}
    \addplot    +[tmtgpu] table[x=lin_size, y=elapsed_total] \mrtmtgpu; \addlegendentry{tmt-sycl, GPU}
\end{axis}
\end{tikzpicture}
    \caption{Running time comparison of tmt-sycl and \vtkm, on \magrec dataset,
             varying input size.}
\label{fig:mr_comp}
\end{figure}

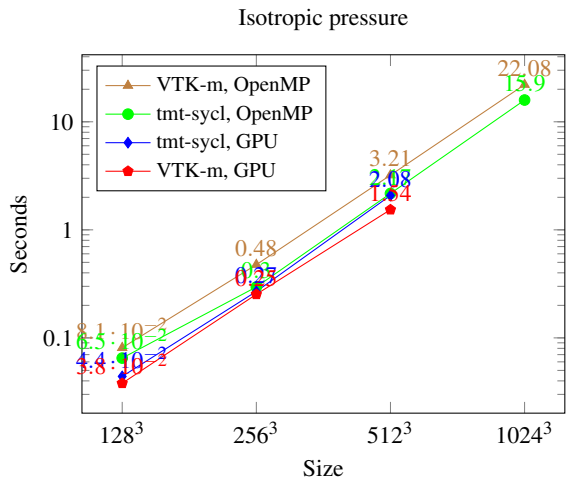
\begin{figure}[t]
\centering
\pgfplotstableread[col sep = semicolon]{data/isotropic_pressure/ip_tmt_sycl_gpu.txt}\iptmtgpu
\pgfplotstableread[col sep = semicolon]{data/isotropic_pressure/ip_tmt_sycl_openmp.txt}\iptmtopenmp
\pgfplotstableread[col sep = semicolon]{data/isotropic_pressure/ip_vtk_gpu.txt}\ipvtkgpu
\pgfplotstableread[col sep = semicolon]{data/isotropic_pressure/ip_vtk_openmp.txt}\ipvtkopenmp
\begin{tikzpicture}
\begin{axis}[
        title={Isotropic pressure},
                ymode=log,
                xmode=log,
                height=2.5in,
                width=.45\textwidth,
                log ticks with fixed point,
                xtick={128,256,512,1024},
                xticklabels={$128^3$, $256^3$,$512^3$,$1024^3$},
                xtick={128,256,512,1024},
                xlabel={Size},
                ylabel={Seconds},
                nodes near coords,      
                point meta=rawy,        
                legend pos=north west,
                legend style={font=\footnotesize},
                legend cell align={left},
                legend columns=1
            ]
    \addplot    +[vtkomp] table[x=lin_size, y=elapsed] \ipvtkopenmp; \addlegendentry{VTK-m, OpenMP}
    \addplot    +[tmtomp] table[x=lin_size, y=elapsed_total] \iptmtopenmp; \addlegendentry{tmt-sycl, OpenMP}
    \addplot    +[tmtgpu] table[x=lin_size, y=elapsed_total] \iptmtgpu; \addlegendentry{tmt-sycl, GPU}
    \addplot    +[vtkgpu] table[x=lin_size, y=elapsed] \ipvtkgpu; \addlegendentry{VTK-m, GPU}
\end{axis}
\end{tikzpicture}
    \caption{Running time comparison of tmt-sycl and \vtkm, on \isopres dataset,
             varying input size.}
\label{fig:ip_comp}
\end{figure}

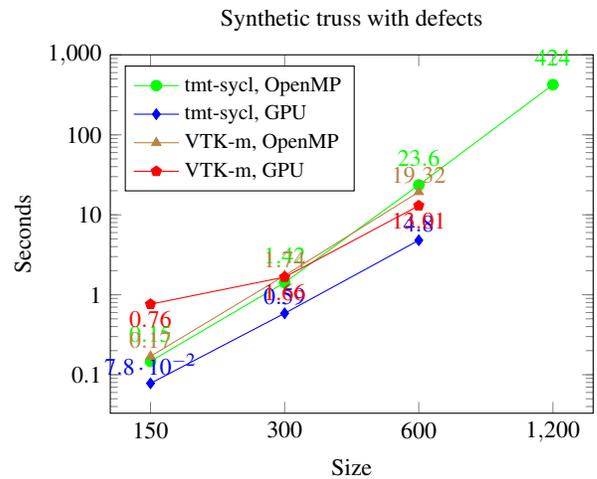
\begin{figure}[t]
\centering
\pgfplotstableread[col sep = semicolon]{data/truss/truss_tmt_sycl_gpu.txt}\trusstmtgpu
\pgfplotstableread[col sep = semicolon]{data/truss/truss_tmt_sycl_openmp.txt}\trusstmtopenmp
\pgfplotstableread[col sep = semicolon]{data/truss/truss_vtk_gpu.txt}\trussvtkgpu
\pgfplotstableread[col sep = semicolon]{data/truss/truss_vtk_openmp.txt}\trussvtkopenmp
\begin{tikzpicture}
\begin{axis}[
        title={Synthetic truss with defects},
                ymode=log,
                xmode=log,
                height=2.5in,
                width=.45\textwidth,
                log ticks with fixed point,
                xtick={150,300,600,1200},
                xlabel={Size},
                ylabel={Seconds},
                legend pos=north west,
                nodes near coords,      
                point meta=rawy,        
                legend style={font=\footnotesize},
                legend cell align={left},
                legend columns=1
            ]
    \addplot    +[tmtomp,
                  every node near coord/.append style={yshift=1ex}]
                    table[x=lin_size, y=elapsed_total] \trusstmtopenmp; \addlegendentry{tmt-sycl, OpenMP}
    \addplot    +[tmtgpu] table[x=lin_size, y=elapsed_total] \trusstmtgpu; \addlegendentry{tmt-sycl, GPU}
    \addplot    +[vtkomp] table[x=lin_size, y=elapsed] \trussvtkopenmp; \addlegendentry{VTK-m, OpenMP}
    \addplot    +[vtkgpu,
                  every node near coord/.append style={yshift=-3ex}]
                    table[x=lin_size, y=elapsed] \trussvtkgpu; \addlegendentry{VTK-m, GPU}
\end{axis}
\end{tikzpicture}
    \caption{Running time comparison of tmt-sycl and \vtkm, on \truss dataset,
             varying input size.}
\label{fig:truss_comp}
\end{figure}

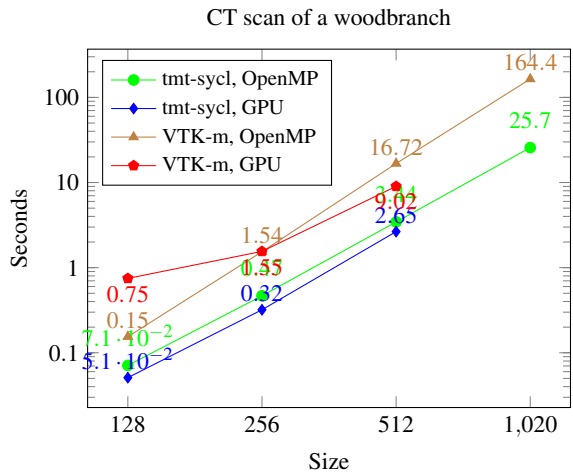
\begin{figure}[t]
\centering
\pgfplotstableread[col sep = semicolon]{data/woodbranch/woodbranch_tmt_sycl_gpu.txt}\woodbranchtmtgpu
\pgfplotstableread[col sep = semicolon]{data/woodbranch/woodbranch_tmt_sycl_openmp.txt}\woodbranchtmtopenmp
\pgfplotstableread[col sep = semicolon]{data/woodbranch/woodbranch_vtk_gpu.txt}\woodbranchvtkgpu
\pgfplotstableread[col sep = semicolon]{data/woodbranch/woodbranch_vtk_openmp.txt}\woodbranchvtkopenmp
\begin{tikzpicture}
\begin{axis}[
        title={CT scan of a woodbranch},
                ymode=log,
                xmode=log,
                height=2.5in,
                width=.45\textwidth,
                log ticks with fixed point,
                xtick={128,256,512,1024},
                xlabel={Size},
                ylabel={Seconds},
                legend pos=north west,
                nodes near coords,      
                point meta=rawy,        
                legend style={font=\footnotesize},
                legend cell align={left},
                legend columns=1
            ]
    \addplot    +[tmtomp,
                  every node near coord/.append style={yshift=1ex}]
                    table[x=lin_size, y=elapsed_total] \woodbranchtmtopenmp; \addlegendentry{tmt-sycl, OpenMP}
    \addplot    +[tmtgpu] table[x=lin_size, y=elapsed_total] \woodbranchtmtgpu; \addlegendentry{tmt-sycl, GPU}
    \addplot    +[vtkomp] table[x=lin_size, y=elapsed] \woodbranchvtkopenmp; \addlegendentry{VTK-m, OpenMP}
    \addplot    +[vtkgpu,
                  every node near coord/.append style={yshift=-3ex}]
                    table[x=lin_size, y=elapsed] \woodbranchvtkgpu; \addlegendentry{VTK-m, GPU}
\end{axis}
\end{tikzpicture}
    \caption{Running time comparison of tmt-sycl and \vtkm, on \woodbranch dataset,
             varying input size.}
\label{fig:woodbranch_comp}
\end{figure}

\paragraph{Results.} On two datasets, \nyx and \magrec,
our implementation clearly outperforms \vtkm, see
\cref{fig:cosmo_z2_comp,fig:mr_comp}.
On CT scans (\cham, \woodbranch, \truss) our
GPU implementation also shows best running times,
but here the advantage is less visible, see \cref{fig:cham_comp,fig:woodbranch_comp,fig:truss_comp}.
Both algorithms are very data-dependent, and this behavior is not at all
universal: for the \isopres dataset, our implementation is just a bit slower, as shown
in \cref{fig:ip_comp},
but for the \richt dataset, it performs significantly worse, see \cref{fig:rm_comp}.
On the other hand, for the largest variant of \paw, performance
of \vtkm deteriorates, see \cref{fig:pawpawsaurus_comp}. 
One possible explanation for this is that the topology of \nyx and \magrec datasets is
richer: the number of branches in the merge tree for $512^3$ samples
from cosmological and magnetic reconnection datasets
is between $7\cdot 10^6$ and $8 \cdot 10^6$, while for \isopres
the tree consists of
much fewer branches, around $5 \cdot 10^5$.
Thus, if the function is expected to be topologically complex,
the triplet merge tree algorithm has an advantage.

\begin{figure*}
\begin{subfigure}[h]{0.45\linewidth}
\includegraphics[width=\linewidth]{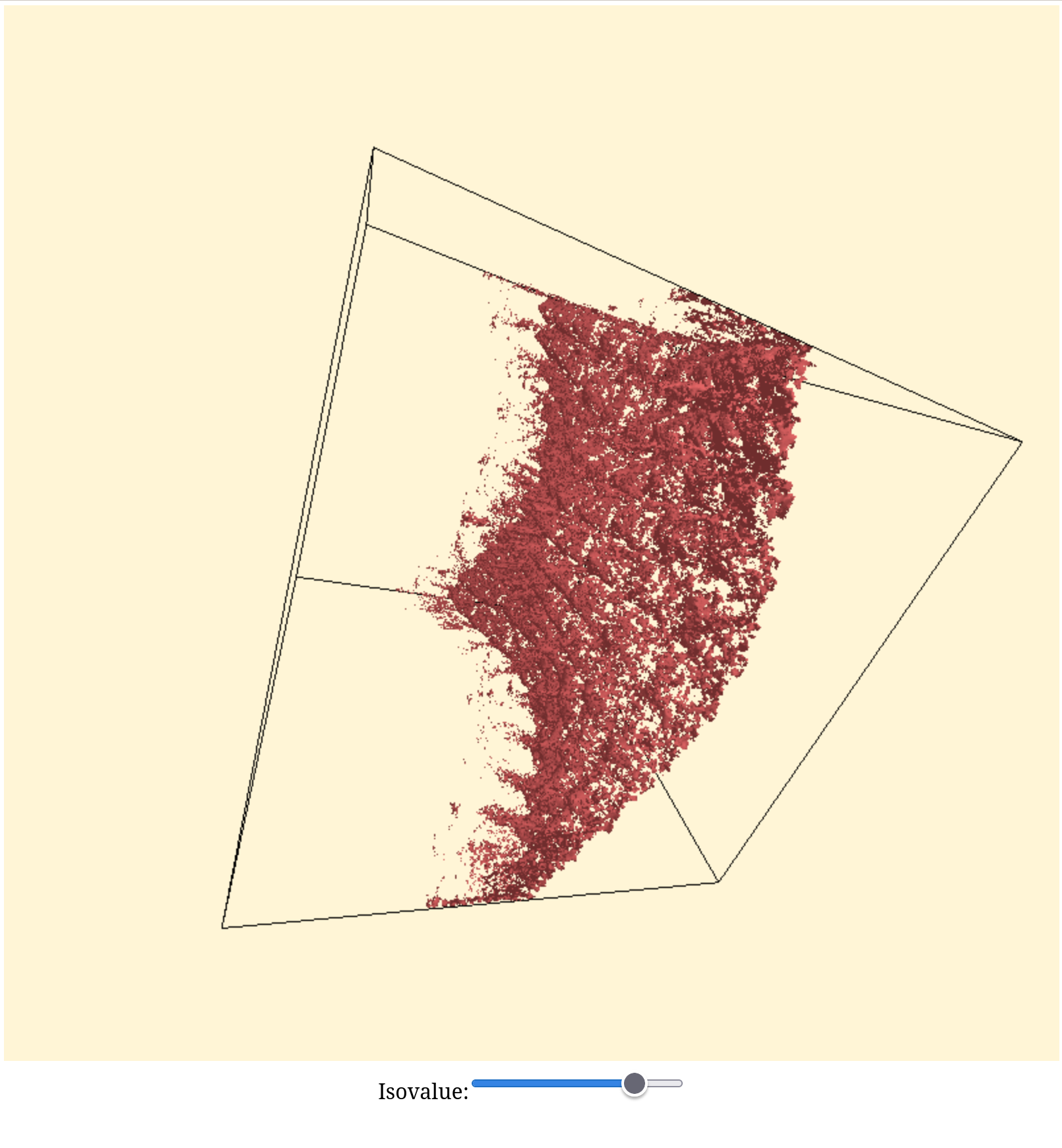}
\caption{Slightly above the threshold}
\end{subfigure}
\hfill
\begin{subfigure}[h]{0.45\linewidth}
\includegraphics[width=\linewidth]{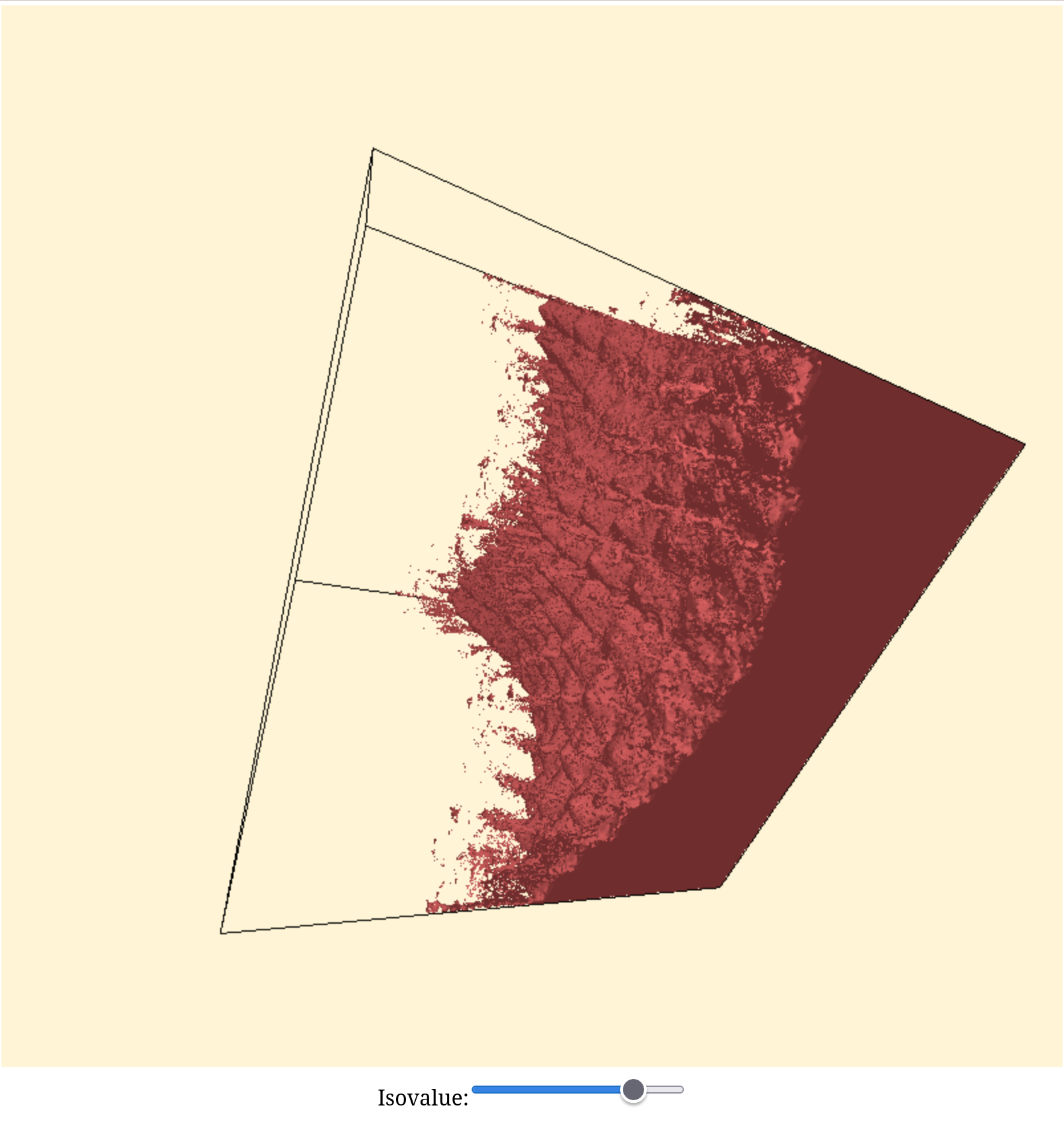}
\caption{Slightly below the threshold}
\end{subfigure}%
\caption{Screenshots of visualization of \richt dataset for two isovalues.}
\label{fig:rm_visual}
\end{figure*}

We note that the topological complexity is not the only factor.
The triplet merge tree algorithm's worst-case complexity is quadratic, and it is
likely that Richtmyer--Meshkov data set is triggering this behavior.
This data set has another distinctive feature. Using
the online visualization tool \cite{klacdb},
we see that at first the evolution of the superlevel sets
is similar to CT scans: there are multiple peaks
where the function value is high and, as we decrease the threshold,
they gradually merge together, so that the volume of the superlevel
set changes slowly. However, then there are two sharp spikes: as we decrease the threshold
by a tiny amount, half of the domain is added to the superlevel set.
One example of this is shown in \cref{fig:rm_visual}.

We summarize all comparisons in \cref{tbl:speedups}.
When we compare the GPU versions of our implementation and \vtkm,
we often see a speedup by a factor of 10--14
for the smallest versions of the datasets, $128^3$ and similar sizes.
This likely has to do with some initial setup computation performed by the
\vtkm code rather than with the difference between the algorithms.
For larger sizes, we often outperform \vtkm by a factor that ranges between $1.5$
and $4.8$.

Perhaps the most disappointing result of our experiments is that we never see a
significant improvement when switching from OpenMP parallelism to GPU; in fact,
sometimes the latter is slightly slower.
The highest speed up that we observed was for \truss dataset, where our GPU
implementation performed almost 5 times faster than its OpenMP variant, but this is an exception.
Typically, we gain between 10 to 50 percent speed up when we move to GPU.
The difference is too small to be seen in the plots with logarithmic scale.
As mentioned in the footnote, we cannot fully attribute this to the time
needed to transfer data to and from the device.
Of course, the ability to compute the tree directly on the GPU is still useful:
either to free the CPU for another task, or especially, when the input data
already resides on the device (e.g., as part of a numerical simulation or
machine learning pipeline).

Finally, we evaluate the strong scaling of the OpenMP versions of the codes, as we increase the number of threads
processing the datasets of size $512^3$,
as shown in \cref{fig:scaling_tmt_vtkm_omp,fig:scaling_tmt_vtkm_omp_rm}\footnote{This experiment
was performed on a machine that has not one, but two CPUs with 20 physical cores each.
Other characteristics of the two machines are the same.}.
As we can see, our implementation usually scales better than \vtkm,
showing some improvement even as we go from 64 to 80 threads,
even though at this number of threads we are using hyperthreading (there are only
40 physical cores available).


\begin{table*}[]
\ra{1.2}
\begin{tabular}{@{}llrrrr@{}}\toprule
Dataset      & Size                         & Comparison OpenMP  & Comparison GPU & TMT-SYCL GPU speedup & \vtkm GPU speedup \\
\hline
\woodbranch  & $128 \times 128 \times 128$  & 2.177                     & 14.659                & 1.392           & 0.207                                                         \\
             & $256 \times 256 \times 256$  & 3.294                     & 4.853                 & 1.459           & 0.991                                                         \\
             & $512 \times 512 \times 512$  & 4.859                     & 3.404                 & 1.298           & 1.853                                                         \\
\hline
\cham        & $128 \times 128 \times 128$  & 1.602                     & 13.091                & 1.526           & 0.187                                                         \\
             & $256 \times 256 \times 256$  & 0.966                     & 3.068                 & 1.578           & 0.496                                                         \\
             & $512 \times 512 \times 512$  & 2.468                     & 1.504                 & 1.046           & 1.717                                                         \\
\hline
\truss        & $128 \times 128 \times 128$ & 1.155                     & 9.770                 & 1.885           & 0.223                                                         \\
             & $256 \times 256 \times 256$  & 1.223                     & 2.838                 & 2.423           & 1.045                                                         \\
             & $512 \times 512 \times 512$  & 0.818                     & 2.710                 & 4.917           & 1.485                                                         \\
\hline
\magrec      & $128 \times 128 \times 128$  & 72.489                    & 14.362                & 0.957           & 4.833                                                         \\
             & $256 \times 256 \times 256$  & 13.953                    & 4.278                 & 1.070           & 3.491                                                         \\
             & $512 \times 512 \times 512$  & 6.389                     & 2.997                 & 1.073           & 2.288                                                         \\
\hline
\isopres     & $128 \times 128 \times 128$  & 1.688                     & 0.905                 & 1.143           & 2.132                                                         \\
             & $256 \times 256 \times 256$  & 1.694                     & 0.984                 & 1.093           & 1.881                                                         \\
             & $512 \times 512 \times 512$  & 1.506                     & 0.765                 & 1.060           & 2.087                                                         \\
\hline
\paw         & $119 \times 80 \times 136$   & 2.022                     & 18.417                & 1.282           & 0.141                                                         \\
             & $238 \times 160 \times 272$  & 2.053                     & 5.113                 & 1.318           & 0.529                                                         \\
             & $476 \times 320 \times 544$  & 2.803                     & 2.261                 & 1.411           & 1.750                                                         \\
             & $952 \times 640 \times 1088$ & 4.791                     & 83.137                & 1.135           & 0.065                                                         \\
\hline
\jicf        & $176 \times 135 \times 137$  & 2.294                     & 11.631                & 1.386           & 0.273                                                         \\
             & $352 \times 270 \times 275$  & 2.075                     & 2.745                 & 1.196           & 0.904                                                         \\
             & $704\times 540 \times 550$   & 1.476                     & 1.287                 & 1.144           & 1.312                                                         \\
\hline
\richt       & $128 \times 128 \times 128$  & 0.772                     & 0.390                 & 0.920           & 1.821                                                         \\
             & $256 \times 256 \times 256$  & 0.290                     & 0.313                 & 1.878           & 1.735                                                         \\
             & $512 \times 512 \times 512$  & 0.023                     & 0.218                 & 19.520          & 2.047                                                         \\
\hline
\nyx         & $128 \times 128 \times 128$  & 3.068                     & 1.852                 & 0.966           & 1.600                                                         \\
             & $256 \times 256 \times 256$  & 3.777                     & 2.191                 & 0.986           & 1.701                                                         \\
             & $512 \times 512 \times 512$  & 4.042                     & 4.412                 & 0.915           & 0.838                                                         \\
\bottomrule
\end{tabular}
\caption{Comparison of our implementation to \vtkm.  Comparison columns for OpenMP and GPU: running time of \vtkm divided
    by the running time of TMT-SYCL (greater than $1$ means our implementation is faster). TMT-SYCL GPU speed-up: our running time on
    OpenMP divided by the running time on GPU (speedup factor of switching to GPU). \vtkm GPU speed-up: \vtkm running time on
    OpenMP divided by \vtkm running time on GPU (speed-up factor of switching to GPU for \vtkm).
    }
\label{tbl:speedups}
\end{table*}

\begin{figure}[t]
\centering
\pgfplotstableread[col sep = semicolon]{data/richtmyer_meshkov/rm_tmt_sycl_gpu.txt}\rmtmtgpu
\pgfplotstableread[col sep = semicolon]{data/richtmyer_meshkov/rm_tmt_sycl_openmp.txt}\rmtmtopenmp
\pgfplotstableread[col sep = semicolon]{data/richtmyer_meshkov/rm_vtk_gpu.txt}\rmvtkgpu
\pgfplotstableread[col sep = semicolon]{data/richtmyer_meshkov/rm_vtk_openmp.txt}\rmvtkopenmp
\begin{tikzpicture}
\begin{axis}[
        title={Richtmyer--Meshkov},
                ymode=log,
                xmode=log,
                height=2.5in,
                width=.45\textwidth,
                log ticks with fixed point,
                xtick={128,256,512,1024},
                xticklabels={$128^3$,$256^3$,$512^3$,$1024^3$},
                xlabel={Size},
                ylabel={Seconds},
                legend pos=north west,
                nodes near coords,      
                point meta=rawy,        
                legend style={font=\footnotesize},
                legend cell align={left},
                legend columns=1
            ]
    \addplot    +[tmtomp,
                  every node near coord/.append style={yshift=1ex}]
                    table[x=lin_size, y=elapsed_total] \rmtmtopenmp; \addlegendentry{tmt-sycl, OpenMP}
    \addplot    +[tmtgpu] table[x=lin_size, y=elapsed_total] \rmtmtgpu; \addlegendentry{tmt-sycl, GPU}
    \addplot    +[vtkomp] table[x=lin_size, y=elapsed] \rmvtkopenmp; \addlegendentry{VTK-m, OpenMP}
    \addplot    +[vtkgpu,
                  every node near coord/.append style={yshift=-3ex}]
                    table[x=lin_size, y=elapsed] \rmvtkgpu; \addlegendentry{VTK-m, GPU}
\end{axis}
\end{tikzpicture}
    \caption{Running time comparison of tmt-sycl and \vtkm, on \richt dataset,
             varying input size.}
\label{fig:rm_comp}
\end{figure}

\begin{figure}[t]
\centering
\pgfplotstableread[col sep = semicolon]{data/chameleon/cham_tmt_sycl_gpu.txt}\chamtmtgpu
\pgfplotstableread[col sep = semicolon]{data/chameleon/cham_tmt_sycl_openmp.txt}\chamtmtopenmp
\pgfplotstableread[col sep = semicolon]{data/chameleon/cham_vtk_gpu.txt}\chamvtkgpu
\pgfplotstableread[col sep = semicolon]{data/chameleon/cham_vtk_openmp.txt}\chamvtkopenmp
\begin{tikzpicture}
\begin{axis}[
        title={Chameleon CT Scan},
                ymode=log,
                xmode=log,
                height=2.5in,
                width=.45\textwidth,
                log ticks with fixed point,
                xtick={128,256,512,1024},
                xticklabels={$128$,$256$,$512$,$1024$},
                xlabel={Size},
                ylabel={Seconds},
                legend pos=north west,
                nodes near coords,      
                point meta=rawy,        
                legend style={font=\footnotesize},
                legend cell align={left},
                legend columns=1
            ]
    \addplot    +[tmtomp,
                  every node near coord/.append style={yshift=1ex}]
                    table[x=lin_size, y=elapsed_total] \chamtmtopenmp; \addlegendentry{tmt-sycl, OpenMP}
    \addplot    +[tmtgpu] table[x=lin_size, y=elapsed_total] \chamtmtgpu; \addlegendentry{tmt-sycl, GPU}
    \addplot    +[vtkomp] table[x=lin_size, y=elapsed] \chamvtkopenmp; \addlegendentry{VTK-m, OpenMP}
    \addplot    +[vtkgpu,
                  every node near coord/.append style={yshift=-3ex}]
                    table[x=lin_size, y=elapsed] \chamvtkgpu; \addlegendentry{VTK-m, GPU}
\end{axis}
\end{tikzpicture}
    \caption{Running time comparison of tmt-sycl and \vtkm, on \cham dataset,
             varying input size.}
\label{fig:cham_comp}
\end{figure}

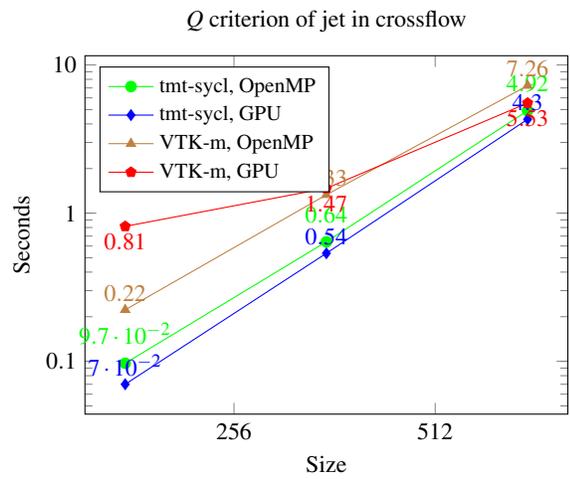
\begin{figure}[t]
\centering
\pgfplotstableread[col sep = semicolon]{data/jicf/jicf_tmt_sycl_gpu.txt}\jicftmtgpu
\pgfplotstableread[col sep = semicolon]{data/jicf/jicf_tmt_sycl_openmp.txt}\jicftmtopenmp
\pgfplotstableread[col sep = semicolon]{data/jicf/jicf_vtk_gpu.txt}\jicfvtkgpu
\pgfplotstableread[col sep = semicolon]{data/jicf/jicf_vtk_openmp.txt}\jicfvtkopenmp
\begin{tikzpicture}
\begin{axis}[
        title={$Q$ criterion of jet in crossflow},
                ymode=log,
                xmode=log,
                height=2.5in,
                width=.45\textwidth,
                log ticks with fixed point,
                xtick={128,256,512,1024},
                xlabel={Size},
                ylabel={Seconds},
                legend pos=north west,
                nodes near coords,      
                point meta=rawy,        
                legend style={font=\footnotesize},
                legend cell align={left},
                legend columns=1
            ]
    \addplot    +[tmtomp,
                  every node near coord/.append style={yshift=1ex}]
                    table[x=lin_size, y=elapsed_total] \jicftmtopenmp; \addlegendentry{tmt-sycl, OpenMP}
    \addplot    +[tmtgpu] table[x=lin_size, y=elapsed_total] \jicftmtgpu; \addlegendentry{tmt-sycl, GPU}
    \addplot    +[vtkomp] table[x=lin_size, y=elapsed] \jicfvtkopenmp; \addlegendentry{VTK-m, OpenMP}
    \addplot    +[vtkgpu,
                  every node near coord/.append style={yshift=-3ex}]
                    table[x=lin_size, y=elapsed] \jicfvtkgpu; \addlegendentry{VTK-m, GPU}
\end{axis}
\end{tikzpicture}
    \caption{Running time comparison of tmt-sycl and \vtkm, on \jicf dataset,
             varying input size.}
\label{fig:jicf_comp}
\end{figure}

\begin{figure}[t]
\centering
\pgfplotstableread[col sep = semicolon]{data/pawpawsaurus/paw_tmt_sycl_gpu.txt}\pawpawsaurustmtgpu
\pgfplotstableread[col sep = semicolon]{data/pawpawsaurus/paw_tmt_sycl_openmp.txt}\pawpawsaurustmtopenmp
\pgfplotstableread[col sep = semicolon]{data/pawpawsaurus/paw_vtk_gpu.txt}\pawpawsaurusvtkgpu
\pgfplotstableread[col sep = semicolon]{data/pawpawsaurus/paw_vtk_openmp.txt}\pawpawsaurusvtkopenmp
\begin{tikzpicture}
\begin{axis}[
        title={CT Scan of Pawpawsaurus},
                ymode=log,
                xmode=log,
                height=2.5in,
                width=.45\textwidth,
                log ticks with fixed point,
                xtick={128,256,512,1024},
                xlabel={Size},
                ylabel={Seconds},
                legend pos=north west,
                nodes near coords,      
                point meta=rawy,        
                legend style={font=\footnotesize},
                legend cell align={left},
                legend columns=1
            ]
    \addplot    +[tmtomp,
                  every node near coord/.append style={yshift=1ex}]
                    table[x=lin_size, y=elapsed_total] \pawpawsaurustmtopenmp; \addlegendentry{tmt-sycl, OpenMP}
    \addplot    +[tmtgpu] table[x=lin_size, y=elapsed_total] \pawpawsaurustmtgpu; \addlegendentry{tmt-sycl, GPU}
    \addplot    +[vtkomp] table[x=lin_size, y=elapsed] \pawpawsaurusvtkopenmp; \addlegendentry{VTK-m, OpenMP}
    \addplot    +[vtkgpu,
                  every node near coord/.append style={yshift=-3ex}]
                    table[x=lin_size, y=elapsed] \pawpawsaurusvtkgpu; \addlegendentry{VTK-m, GPU}
\end{axis}
\end{tikzpicture}
    \caption{Running time comparison of tmt-sycl and \vtkm, on \paw dataset,
             varying input size.}
\label{fig:pawpawsaurus_comp}
\end{figure}
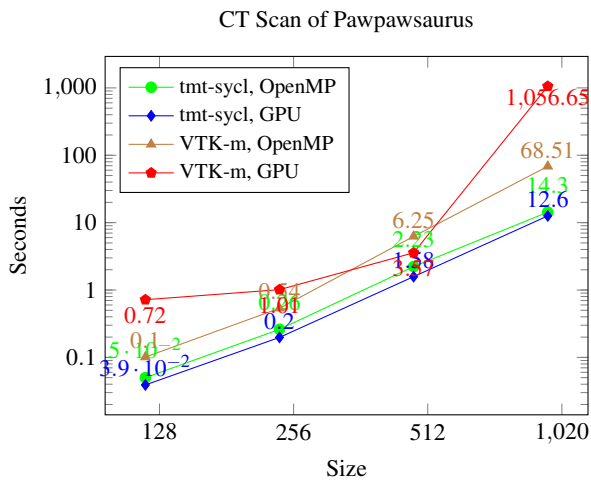

\begin{figure}[t]
\centering

\pgfplotstableread[col sep = semicolon]{data/scaling_z2_cole.txt}\cosmoscaletmt
\pgfplotstableread[col sep = semicolon]{data/scaling_z2_vtkm_cole.txt}\cosmoscalevtk
\pgfplotstableread[col sep = semicolon]{data/scaling_mr_cole.txt}\mrscaletmt
\pgfplotstableread[col sep = semicolon]{data/scaling_mr_vtkm_cole.txt}\mrscalevtk
\pgfplotstableread[col sep = semicolon]{data/scaling_rm_cole.txt}\rmscaletmt
\pgfplotstableread[col sep = semicolon]{data/scaling_rm_vtkm_cole.txt}\rmscalevtk
\pgfplotstableread[col sep = semicolon]{data/scaling_ip_cole.txt}\ipscaletmt
\pgfplotstableread[col sep = semicolon]{data/scaling_ip_vtkm_cole.txt}\ipscalevtk
\begin{tikzpicture}
\begin{axis}[
        title={Strong scaling},
                ymode=log,
                xmode=log,
                ymax=500,
                height=2.5in,
                width=.45\textwidth,
                log ticks with fixed point,
                xtick={1,2,4,8,16,40,80},
                xlabel={threads},
                ylabel={Seconds},
                legend pos=north east,
                legend cell align={left},
                legend columns=3,
                transpose legend,
                point meta=rawy,        
                legend style={font=\footnotesize},
                legend style={fill=none},
            ]
    \addplot    +[tmtompcosmo] table[x=threads, y=elapsed_total] \cosmoscaletmt; \addlegendentry{tmt-sycl, NYX}
    \addplot    +[tmtompip,
                  nodes near coords,
                  every node near coord/.append style={yshift=-3ex}]
                            table[x=threads, y=elapsed_total] \ipscaletmt; \addlegendentry{tmt-sycl, IP}
    \addplot    +[tmtompmr] table[x=threads, y=elapsed_total] \mrscaletmt; \addlegendentry{tmt-sycl, MAG}

    \addplot    +[vtkompcosmo] table[x=threads, y=elapsed_total] \cosmoscalevtk; \addlegendentry{VTK-m, NYX}
    \addplot    +[vtkompip] table[x=threads, y=elapsed_total] \ipscalevtk; \addlegendentry{VTK-m, IP}
    \addplot    +[vtkompmr, nodes near coords] table[x=threads, y=elapsed_total] \mrscalevtk; \addlegendentry{VTK-m, MAG}
\end{axis}
\end{tikzpicture}
    \caption{Strong scaling of OpenMP version of tmt-sycl and \vtkm.}
\label{fig:scaling_tmt_vtkm_omp}
\end{figure}

\begin{figure}[t]
\centering

\pgfplotstableread[col sep = semicolon]{data/scaling_rm_cole.txt}\rmscaletmt
\pgfplotstableread[col sep = semicolon]{data/scaling_rm_vtkm_cole.txt}\rmscalevtk
\begin{tikzpicture}
\begin{axis}[
        title={Strong scaling, Richtmyer--Meshkov},
                ymode=log,
                xmode=log,
                height=2.5in,
                width=.45\textwidth,
                log ticks with fixed point,
                xtick={1,2,4,8,16,40,80},
                xlabel={threads},
                ylabel={Seconds},
                nodes near coords,      
                point meta=rawy,        
                legend pos=north east,
                legend style={font=\footnotesize},
                legend cell align={left},
                legend columns=1
            ]
    \addplot    +[tmtomprm] table[x=threads, y=elapsed_total] \rmscaletmt; \addlegendentry{tmt-sycl, RM}
    \addplot    +[vtkomprm] table[x=threads, y=elapsed_total] \rmscalevtk; \addlegendentry{VTK-m, RM}
\end{axis}
\end{tikzpicture}
    \caption{Strong scaling of OpenMP version of tmt-sycl and \vtkm.}
\label{fig:scaling_tmt_vtkm_omp_rm}
\end{figure}

\section{Conclusion}
We presented a GPU implementation of a merge tree computation that often performs better than the previously available implementation in \vtkm
for inputs that are topologically rich and do not undergo drastic changes in the volume
of the sublevel set. We believe that such inputs are reasonably common, so
it makes sense to use the GPU version of the triplet merge algorithm.
There are two questions for future work.
First, how can we gain more from GPU parallelism?
Second, how to adapt our algorithm to a distributed setting?
This is not trivial because of the $32$ bit limitation
that we had to make in order to pack an edge into $64$ bits.
While innocent for a single GPU, for large-scale simulations
that run on the GPUs of multiple nodes, it becomes a problem.

\section*{Acknowledgements}
This work was supported by the U.S. Department of Energy, Office of Science, Office of Advanced Scientific Computing Research, Scientific Discovery through Advanced Computing (SciDAC) program, under Contract Number DE-AC02-05CH11231 at
Lawrence Berkeley National Laboratory.

\bibliographystyle{abbrv-doi}
\bibliography{references}

\end{document}